\documentclass{article}
\usepackage{graphicx}% Include figure files
\usepackage{dcolumn}% Align table columns on decimal point
\usepackage{bm}% bold math
\usepackage{color} %color
\usepackage{amsmath}
\usepackage{amssymb}
\usepackage[font={footnotesize,it}]{caption}

\usepackage{titlesec}

\titleformat*{\section}{\large\bfseries}
\titleformat*{\subsection}{\normalsize\bfseries}
\titleformat*{\subsubsection}{\small\bfseries}

\begin{document}

\pagestyle{empty}
\begin{center}
%\vskip 1.4 cm
{\LARGE{\bf Optimized Photonic Gauge of Extreme High Vacuum with Petawatt Lasers}}
\end{center}
\vskip 10pt
\begin{center}
{\large 
\'Angel Paredes$^1$, David Novoa$^2$, Daniele Tommasini$^1$ and H\'ector Mas$^1$}\\
\end{center}
\vskip 10pt
\begin{center}
\textit{$^1$ Departamento de F\'\i sica Aplicada,
Universidade de Vigo, As Lagoas s/n, Ourense, ES-32004 Spain;\\
$^2$ Max Planck Institute for the Science of Light, G\"unther-Scharowsky Str. 1, 91058 Erlangen, Germany.
}\\
\end{center}
\vskip.2cm
{\small Corresponding author: {\tt angel.paredes@uvigo.es}}

\vspace{15pt}

\begin{center}
\textbf{Abstract}
\end{center}

\vspace{4pt}{\small \noindent 
One of the latest proposed applications of ultra-intense laser pulses is their possible use to gauge extreme high vacuum by measuring the photon radiation resulting from nonlinear Thomson scattering within a vacuum tube. Here, we provide a complete analysis of the process, computing the expected rates and spectra both for linear and circular polarizations of the laser pulses, taking into account the effect of the time envelope in a slowly-varying envelope approximation. 
We also design a realistic experimental configuration allowing for the implementation of 
the idea and compute the corresponding geometric efficiencies. Finally, we develop an optimization procedure for this Photonic Gauge of Extreme High Vacuum at 
high repetition rate Petawatt and multi-Petawatt laser facilities, such as VEGA, JuSPARC and ELI.
}

\vfill

\newpage

\setcounter{page}{1}
\pagestyle{plain}

%-----------------------------------------------------------------------------

%\pacs{42.62.-b, 07.30.Dz, 41.60.-m, 52.38.-r}

%--------------------------------------------------------------------------------

%\twocolumn
\footnotesize

\section{Introduction}

Since the invention of chirped pulse amplification \cite{CPA}, the achievable peak intensity of laser light has increased by more than eight orders of magnitude. The current record intensity, achieved at HERCULES few years ago \cite{HERCULES2008}, is $2\times10^{22}$\ W/cm$^2$, and it may be improved by an order of magnitude by focusing Petawatt (PW) laser pulses close to the diffraction limit. Such enormous intensities are obtained by squeezing the laser pulses both in space and in time, packing a huge number of photons ($\sim10^{20}$ for a PW laser of duration 
30 fs and wavelength $\lambda_0=800$nm) in a volume of the order of a few $\mu$m$^3$.
 
These new sources of radiation have very relevant implications in many fields, such as charged particle acceleration, fast ignition of fusion targets, laboratory simulation of astrophysical conditions and experimental probing of extreme physical regimes \cite{mourou06,qvac-reviews}. In addition, they have been proposed as new tools to test the quantum polarization properties of the vacuum \cite{PPSVsearch} and search for new physics, such as axion-like or mini charged particles \cite{new_physics}.

Recently, it has been suggested \cite{pressure} that they can also be used 
to gauge Extreme High-Vacuum  (XHV)  \cite{redhead98,redhead,calcatelli}, corresponding to pressures $p<10^{-10}$ Pa. 
Having a set-up without the electric field of usual ionization gauges may be useful to circumvent their
limitations---see also \cite{Chen87,haffner} for alternative approaches to XHV gauging. This application of ultra intense lasers to vacuum science is of topical interest since the number and availability of such facilities is expected to increase at a very significant rate in the near future. Moreover, many of the experiments that have been proposed to search vacuum polarization effects and new physics at such facilities, as cited above \cite{PPSVsearch,new_physics}, require the generation and calibration of XHV to control the background noise stemming from the interaction of the laser pulse with the classical vacuum and compute the final sensitivities for the signal. The fact that this can be done using the ultra intense laser itself is a most welcome result. 

The idea behind this proposal is fairly simple:
photons from the laser pulse are scattered by
electrons in the vacuum chamber.
The number of scattered photons is directly proportional to the
electron density and therefore to the pressure. 
Background noise can in principle
be kept below the signal by appropriately synchronizing the measurements to the passage of the pulse through the detection region.
However, even though the physical principle behind the technique is rather straightforward, the key
 question to be answered regarding its viability is whether the photon signal is strong 
enough to be measured. This sets a lower limit for the measurable electron density in a given
facility and with a given photon detection system.
Here, we provide a complete analysis of the process, computing the expected rates and spectra both for linear and circular polarizations of the laser pulses, taking into account the effect of the time envelope in a slowly-varying envelope approximation. 
We also design a realistic experimental configuration allowing for the implementation of 
the idea and compute the corresponding geometric efficiencies. Finally, we develop an optimization procedure for this Photonic Gauge of Extreme High Vacuum at 
high repetition rate Petawatt and multi-Petawatt laser facilities, such as VEGA, JuSPARC and ELI.

The outline of this work is the following: In section \ref{sec II}, we use a slowly-varying envelope approximation (in space and time)
to compute the average number of nonlinear-Thomson-scattered photons by an electron from an intense Gaussian-shaped pulse of light. In particular, we
study  the impact of the use of circular polarization and the  corrections involved after taking into account the time envelope of the pulses. In section \ref{sec: collect}, we take into account that in an eventual XHV measurement,
only a fraction of the scattered photons may be actually measured and therefore discuss the geometric
efficiency in terms of a few simple parameters. This allows to develop an optimization procedure 
for a Photonic Gauge  of XHV. Section \ref{sec: quant} is devoted to give some quantitative estimates
of the possibilities of detection of scattered photons
 at present 
and future PW and multi-PW facilities. Section \ref{sec: discussion} addresses several further questions such as the maximum pressure this photonic gauge might potentially 
handle, the possibility of using table-top high-intensity lasers for the vacuum measurements and the actual spectrum of scattered radiation we might expect in a realistic 
situation. In section \ref{sec: conclusions} we present our conclusions. Some technical details are relegated to two
appendices.

\section{Number of scattered photons per pulse}
\label{sec II}

The dominant interaction of an ultra-intense
beam with an extremely rarefied gas is nonlinear, relativistic, Thomson scattering \cite{NTS}.
In this section, we will use the results of \cite{sarachik} to estimate the number
of scattered photons when a pulse traverses a vacuum chamber
in which we assume there is a uniform number of non-relativistic 
free electrons per unit volume $n_e$.
The pulse will be modelled as a standard Gaussian beam in the transverse direction 
(wavelength $\lambda_0$, beam waist $w_0$, peak intensity $I_0$) 
\begin{equation}
I=I_0\left(\frac{w_0}{w(z)}\right)^2e^{-\frac{2r^2}{w(z)^2}},
\label{Igauss}
\end{equation}
with $w(z)=w_0\sqrt{1+z^2/z_R^2}$, where
the Rayleigh range is $z_R=\pi\,w_0^2/\lambda_0$. 
For simplicity, a sharp time envelope of duration $\tau$, such that the pulse energy is given
by $E_{pulse}=\tau\,I_0\pi\,w_0^2/2$, will be considered.
At the end of the section,  the consequences of non-trivial time envelopes will be discussed.

In the following, most of the equations will be given in terms of 
a dimensionless parameter $q$, related to
the intensity $I$ as
\begin{equation}
q^2 = \frac{2I\,r_0 \lambda_0^2}{\pi \,m_e c^3},
\label{q2}
\end{equation}
where $r_0\approx 2.82\times 10^{-15}$m is the classical electron radius 
and $m_e$ is the electron mass.
$q \approx 1$ signals the onset of relativistic effects, while for $q\ll 1$ linear Thomson scattering is a good approximation. In order to catch a glimpse of realistic values
at present day facilities, let us consider a one PW peak power infrared pulse with
$\lambda_0=800$nm. Taking $w_0=1\mu$m (near to the diffraction limit), we find
$I_0 \approx 0.6 \times 10^{23}$ W/cm$^2$ corresponding to $q_0 \approx 170$ whereas
for $w_0=20\mu$m, the peak intensity is $I_0 \approx 1.6 \times 10^{20}$ W/cm$^2$
and $q_0 \approx 8.6$.

Introducing dimensionless quantities $\rho=r/w_0$ and $\xi=z/z_R$, 
we can write the position-dependent value of $q$ for a Gaussian beam as
\begin{equation}
q^2 (\rho,\xi)= q_0^2 \frac{1}{1+\xi^2}\exp\left(-\frac{2\rho^2}{1+\xi^2}\right).
\label{qpos}
\end{equation}

\subsection{Relativistic Thomson scattering}
\label{sec: RelThom}

The differential cross section for the relativistic
Thomson scattering of plane wave radiation by the electrons of a gas 
has been computed analytically long ago by Sarachik and Schappert \cite{sarachik}. 
 The computation neglects quantum effects,
$n\,h\,c/\lambda_0 \ll m_e c^2$, where $n$ is the harmonic order, and radiation reaction, $q_0^2 \ll \lambda_0 / r_0$, conditions which
are always met at optical frequencies.
The results for plane waves can be used
in a realistic set-up depending on whether a kind of slowly-varying envelope approximation is
sound.  This requires the number of optical periods in the pulse to be large, $\tau \gg \lambda_0/c$,
and the transverse excursion of the electron to be small compared to the beam radius,
i.e. $w_0 \gg q_0 \lambda_0/2\pi$. This latter condition results in
\begin{equation}
w_0 \gg \frac{\lambda_0}{\pi}\left(\frac{E_{pulse}r_0}{m_ec^3\tau}\right)^\frac14.
\label{w0cond}
\end{equation}
In the rest of this subsection,  we review part of the results of \cite{sarachik}
are fix the notation.

In the laboratory frame, the power scattered per unit solid angle in
harmonic $n$ when
a plane wave hits a free electron at rest can be written as
\begin{equation}
\frac{dP^{(n)}}{d\Omega}=\frac{e^2 c}{8\epsilon_0\lambda_0^2}f^{(n)}
\end{equation}
in SI units. The spherical coordinates $\theta\in[0,\pi]$, $\varphi\in[0,2\pi]$ are chosen in such a way that
$\theta=0$ corresponds to forward scattering. The form of $f^{(n)}$ depends on the polarization
of the beam. Hereafter we will analyze linear and circular polarization. To do so, we define 
\begin{equation}
{\cal M}=1+\frac12 q^2\sin^2\left(\frac{\theta}{2}\right)\,.
\end{equation}
For linear polarization, the function  $f^{(n)}$ reads then
\begin{eqnarray}
f_l^{(n)}=\frac{q^2n^2}{{\cal M}^4}
\Bigg[\left( 1-\frac{(1+\frac12 q^2) \cos^2\alpha}{{\cal M}^2} \right)(F_1^n)^2+\nonumber\\
-\frac{q\,\cos\alpha (\cos\theta - \frac12 q^2 \sin^2(\theta/2))}{2{\cal M}^2}
F_1^n F_2^n+\frac{q^2 \sin^2 \theta}{16 {\cal M}^2 }(F_2^n)^2\Bigg],
\label{fn}
\end{eqnarray}
where $\cos\alpha=\sin\theta\,\cos\varphi$,  the polarization axis
corresponds to $\varphi=0,\pi$ and the following functions have been introduced
\begin{eqnarray}
F_s^n=\sum_{l=-\infty}^{+\infty}J_l\left(\frac{n\,q^2\sin^2(\theta/2)}{4{\cal M}}\right)\times\nonumber\\
\left[J_{2l+n+s}\left(\frac{q\,n\,\cos\alpha}{{\cal M}}\right)+J_{2l+n-s}\left(\frac{q\,n\,\cos\alpha}{{\cal M}}\right)\right],\nonumber
\end{eqnarray}
where the $J_l$ are Bessel functions of the first kind.

The result for circular polarization is
\begin{equation}
f_c^{(n)}=\frac{2q^2n^2}{{\cal M}^4}
\Bigg[\frac{2(\cos\theta - \frac{q^2}{2}\sin^2(\theta/2))^2}{q^2\sin^2\theta}
J_n^2(n\,\Theta)+J_n'^2(n\,\Theta)\Bigg],
\label{fncirc}
\end{equation}
where
\begin{equation}
\Theta = \frac{q\,\sin\theta}{\sqrt2\,{\cal M}}.
\end{equation}

In the laboratory frame, the $n$'th-harmonic frequency is not just a multiple of the incident one. Instead, the
following relation holds,
\begin{equation}
\lambda^{(n)}={\cal M}\lambda_0/n.
\label{lambdan}
\end{equation}

All these expressions are valid for electrons initially at rest --- equivalent expressions for
rapid electrons for linear and circular polarization 
 were computed in \cite{salamin}, \cite{salamin2}. We stress that even if the electrons reach
relativistic velocities while the pulse is passing, they remain slow afterwards, since typically
no net energy can be transferred to them \cite{lawson}.

\subsection{Photons scattered per electron from a plane wave}
\label{sec: Gamma}

Regarding Eq. (\ref{lambdan}), the laboratory frame energy of $n$'th harmonic photons is given
by $\frac{h\,c\,n}{\lambda_0{\cal M}}$ and thus depends on the 
intensity of the incident wave and the scattering angle.
The number of  photons scattered per unit time and per unit solid angle is
$\frac{dP^{(n)}}{d\Omega}\frac{{\cal M}\,\lambda_0}{h\,c\,n}$.
If a plane wave of duration $\tau$ impinges on a single electron, the number of scattered 
photons for the $n$'th harmonic is
\begin{equation}
N_{\gamma,pw}^{(n)}=\tau  \frac{e^2}{8\epsilon_0\lambda_0 h}\Gamma^{(n)}(q),
\label{ngp}
\end{equation}
with
\begin{equation}
\Gamma^{(n)}(q)=\frac{1}{n}\int_0^{2\pi}\int_0^\pi f^{(n)} {\cal M} \sin\theta d\theta d\varphi.
\label{GammaDef}
\end{equation}

\begin{figure}[htb]
\centerline{\includegraphics[width=0.6\textwidth]{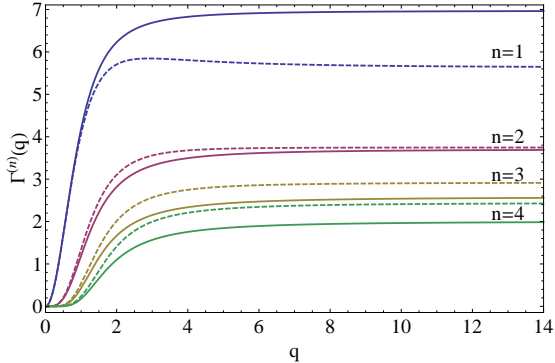}}
\caption{ The function $\Gamma^{(n)}(q)$ found by numerical integration
for linear (solid lines) and circular (dashed lines) polarization, for n=1,\dots,4,
from top to bottom.
}
\label{figGamma}
\end{figure}

For large $q$, all $\Gamma^{(n)}(q)$ tend asymptotically to some constant.
Figure \ref{figGamma} shows the results of the numerical integration
of the expression in Eq. (\ref{GammaDef}) for linear and circular polarization.
Linear polarization produces somewhat more $n=1$ photons when $q$ is 
larger than one, whereas higher harmonics are slightly
enhanced by circular polarization.

For $q\rightarrow 0$, both linear and circular polarizations yield similar values of $\Gamma^{(n)}(q)$. In that limit, the agreement between the curves corresponding to different polarizations is better as $n$ gets reduced 
(see Fig. \ref{figGamma}). This assertion is further confirmed by the lower limit of the analytical representations of $\Gamma^{(n)}(q)$, which are obtained by fitting the curves displayed in Fig. \ref{figGamma} 
to quotients of polynomials (see appendix A).

\subsection{Photons scattered from a Gaussian pulse}
\label{sec: gaussi}

As noted in \cite{pressure}, in order to find the total number of photons
scattered from a realistic pulse, it is crucial to take into account its finite transverse profile.
The previous results for plane waves are useful since, in the spirit of the slowly-varying
envelope approximation introduced in section \ref{sec: RelThom}, a suitable Gaussian profile can be considered, {\it locally}, as plane.

The number of photons scattered from a Gaussian pulse is given
by an integral of the plane wave result over the 
non-trivial profile,
namely $N_\gamma^{(n)} = n_e \int N_{\gamma,pw}^{(n)}(q) d^3 \vec x$, where
$n_e$ is the number of electrons per unit volume and
$N_{\gamma,pw}^{(n)}(q)$ is given in Eq. (\ref{ngp}). The parameter $q$ depends on the
point of space according to Eq. (\ref{qpos}). Using the coordinates $\rho,\xi$
defined at the beginning of this section, we can write:
\begin{equation}
N_\gamma^{(n)}= {\cal K}\int_{-\infty}^\infty\int_0^\infty \rho\,\Gamma^{(n)}(q)   d\rho d\xi,
\label{Ngamma}
\end{equation}
where ${\cal K}$ is a dimensionless quantity,
\begin{equation}
{\cal K}=\frac12 n_e\,c\,\tau\,\pi^2w_0^4\lambda_0^{-2}\alpha,
\label{calK}
\end{equation}
where $\alpha=e^2/4\pi\,\epsilon_0 \hbar c \approx 1/137$ is the fine structure constant.
Notice that $N_\gamma^{(n)}/{\cal K}$ depends only on $n$ and $q_0$, and can be
straightforwardly computed numerically. 

\begin{figure}[htb]
\centerline{\includegraphics[width=0.6\textwidth]{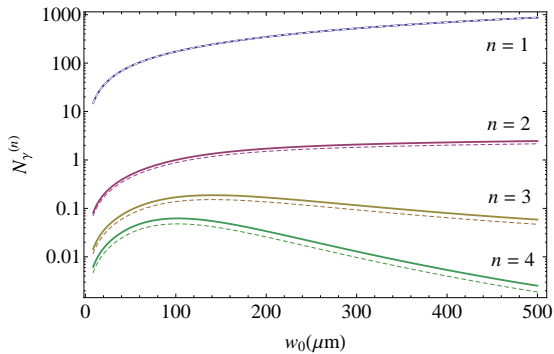}}
\caption{ The number of scattered photons from a Gaussian pulse
as a function of the beam waist. The following parameters have been fixed for the laser
beam
$E_{pulse}=30$J, $\tau=30$fs, $\lambda=800$nm. The pressure has been fixed to
$p=0.5\times 10^{-11}$Pa and the temperature to $T=300K$ such that 
$n_e=2p/k_B T \approx 2.4$mm$^{-3}$, where the factor of 2 comes from considering molecular
hydrogen with two electrons per molecule. The y-axis (x-axis) is given in logarithmic (linear) scale.
}
\label{figNg}
\end{figure}

As shown in \cite{pressure}, $N_\gamma^{(n)}/{\cal K}$ grows as $q_0^3 \sim I_0^\frac32 \sim w_0^{-3}$ 
for large values of the
laser peak intensity. Thus, 
if the remaining parameters are fixed, the number of scattered photons grows linearly with the beam
waist radius $w_0$ (for small enough $w_0$). This behaviour changes for large $w_0$ when the intensities
are low so that harmonic production is suppressed. In fact, just considering the asymptotic
behaviour $\Gamma^{(n)} \propto q^{2n}$ for small values of $q$, we readily find that the integrand
in Eq. (\ref{Ngamma}) is proportional to $q_0^{2n} \propto w_0^{-2n}$. Taking into account
the factor ${\cal K}$, we conclude that $N_\gamma^{(n)} \propto w_0^{4-2n}$ for large $w_0$. This asymptotic dependence holds independently of the chosen polarization.
These rough arguments qualitatively explain the behaviour depicted in Fig.
\ref{figNg}, where a sample numerical computation of the number of scattered photons $N_\gamma^{(n)}$ as a function of the beam waist is shown. In particular, the number of photons produced in the second harmonic $n=2$ tends to a constant value, i.e., they do not display any further dependence on $w_0$ for large waists.
For higher harmonics $n\geq3$, the 
signal drops with increasing $w_0$, as predicted by the analysis developed in this section.
The situation for $n=1$ is subtler since the integral in Eq. (\ref{Ngamma}) diverges in this case.
It is plain
that such result is not physical since the scattering region is always limited. For the plots included in Fig. \ref{figNg},
the integral has been cut at $q=0.01$, the value associated to the barrier suppression regime,
below which the electron of a hydrogen atom cannot
be approximated as free any more (for $\lambda_0=800$nm) \cite{barriersupH}.
As we can appreciate from the figure, $N_\gamma^{(1)}$ grows monotonically with $w_0$.

From the figure, we might be tempted to conclude that the optimal value of $w_0$ for the
measurement of tiny pressures by detecting harmonic-$n$ photons
 would be the one in which $N_\gamma^{(n)}$ reaches its maximum. Nevertheless, this statement
 is naive for at least three reasons. First, it could prove costly or unfeasible to manipulate
 ultra-high power pulses in order to achieve too large waists. Second, the larger the $w_0$, the larger
  the region in which the vacuum gauging is taking place. In a chamber with differential vacuum,
  it would be impossible to measure the vacuum confined in a small region if $w_0$ is too large.
Third, producing more photons does not mean that a larger signal can be measured. If the region
where the scattering is taking place is extensive, it might be impossible to set up an efficient system to collect the emitted photons. We will turn to these questions in section \ref{sec: collect}.

\subsection{Non-trivial time envelope}
\label{sec: envelope}

\begin{figure}[htb]
\centerline{\includegraphics[width=0.6\textwidth]{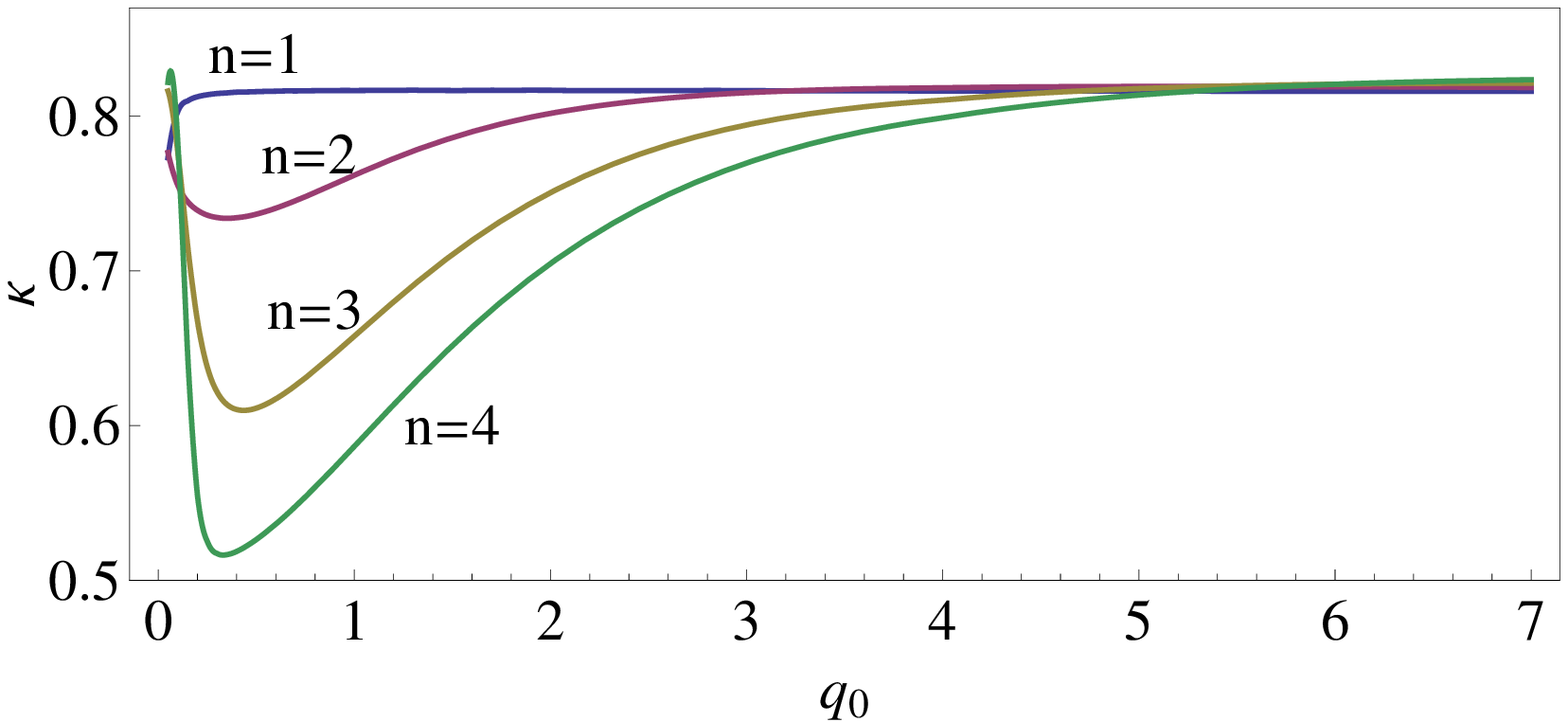}}
\centerline{\includegraphics[width=0.6\textwidth]{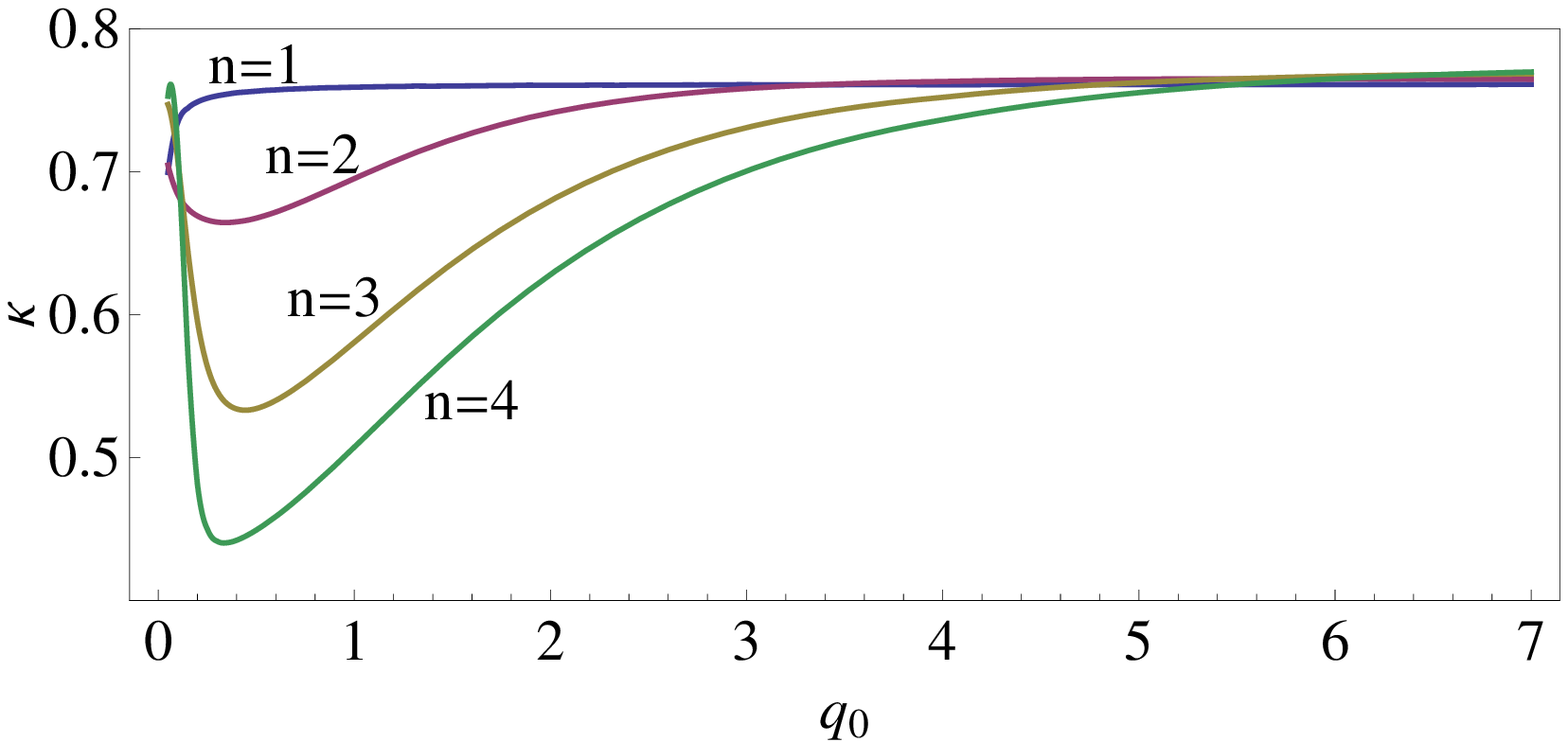}}
\caption{ Multiplicative correction due to non-trivial time envelopes
to the number of scattered photons as a function of $q_0$. On top, the result for the 
Gaussian envelope and below for hyperbolic secant envelope.
}
\label{fig:te}
\end{figure}

Let us  comment 
on the effect of considering a more realistic non-trivial
time envelope instead of a sharp rectangular pulse. We will show that the
number of scattered photons does only 
depend mildly on the envelope and the (simpler) computations of the previous subsections
capture the quantitative results up to a factor of order $1$.
To take into account the time envelope, we describe the spatio-temporal dependent intensity as
$I_{te}=g(\tilde t) I$, where $I$ is given in Eq. (\ref{Igauss}) and $t$ should be understood
as $\tilde t=t-z/c$. If the time envelope varies mildly within a light cycle
($\frac{dg}{d\tilde t}\ll \omega g$), a slowly-varying envelope approximation is valid 
\cite{gibbon} and we may simply use the expression above by including time dependence
in $q$. It is useful to define a time envelope correction parameter as the quotient
\begin{equation}
\kappa\equiv
\frac{N_\gamma^{(n)}|_{te}}{N_\gamma^{(n)}|_{ss}}=\frac
{\int \rho\,\Gamma^{(n)}\left(\sqrt{g(\tilde t)} q(\rho,\xi)\right) d\tilde t\,d\rho\,d\xi}
{\tau\,\int \rho\,\Gamma^{(n)} \left(q(\rho,\xi)\right) d\rho\,d\xi},
\end{equation}
where $N_\gamma^{(n)}|_{ss}$ refers to the computation with a sharp step time envelope,
as in section \ref{sec: gaussi}.
Given the form of the time envelope, $\kappa$ only depends on $q_0$ and $n$. 
As examples, let us consider a Gaussian $g(\tilde t)=e^{-\pi \tilde t^2/\tau^2}$
and a hyperbolic secant $g(\tilde t)=$ sech$(\pi\,\tilde t/\tau)$, chosen
such that $g(0)=1$ and $\int_{-\infty}^\infty g(\tilde t)d\tilde t=\tau$.
In figure \ref{fig:te},  the value of $\kappa$ as a function of $q_0$ is plotted
 for these two time envelopes and for different
 harmonics, considering linear polarization. The plots for circular polarization
 are not shown since they are practically
 coincident with the linear polarization ones. 

The conclusion is that the sharp step
envelope overestimates the number of scattered photons by a factor of order $1$.
The correction factors depend on the shape
of the actual envelope, the harmonic number and the peak intensity, with typical values
around $0.7$ or $0.8$ (see Fig. \ref{fig:te}).

\section{Photon collection and geometric efficiency}
\label{sec: collect}

Up to now, we have computed how many photons are scattered from a given
pulse traversing a vacuum chamber. In this section, we will discuss how
many might be actually measured in a realistic experiment. Since, in any
case, the signal from XHV will be very low, it is essential
to use single-photon detectors, which
 can achieve remarkable quantum efficiencies with state-of-art technology \cite{detectors1,detectors2}. Typically, the size of the 
active region of this kind of detectors is of the order of a few microns.
Then, since the Thomson scattered photons have a wide angular distribution
\cite{pressure}, it is essential to introduce a suitable optical system
in order to have efficient photon collection.

\begin{figure}[htb]
{\centerline  {\includegraphics[width=0.5\columnwidth]{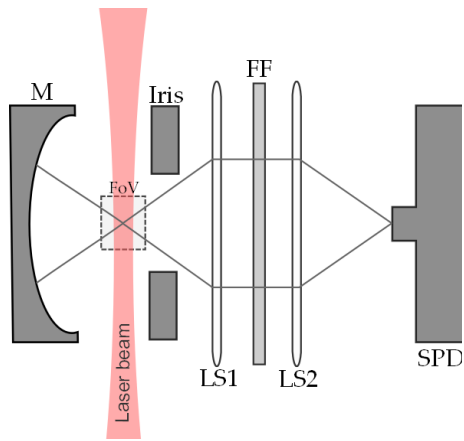}}
} 
\caption{ Sketch of a typical scheme for an efficient photon collection system. M: concave mirror,
FoV: field of view,
 FF: frequency filter, LS1, LS2: lens systems, SPD: single-photon detector.
}
\label{fig:hrs}
\end{figure}

The situation is analogous to the detection of
Hyper-Rayleigh scattering, a well-established technique for the characterization
of nonlinear optical properties of different materials, in particular molecular
hyperpolarizabilities \cite{HRS}: a laser beam traverses the substance to be studied
producing anisotropic faint radiation in a multiple of the incident frequency (usually, incoherent second harmonic) which can be collected and measured
by an optical system which concentrates part of the emitted light into a photomultiplier.
The typical photon collection system --- see for instance \cite{HRS2} ---  is schematically
depicted in Fig. \ref{fig:hrs}. As it can be appreciated in the figure, a parabolic mirror captures the photons that are counter-scattered with respect to the position of the single-photon detector, thus enhancing the 
signal accordingly. An optical system made by filters and lenses allows then to couple most of the scattered light into the detector. A similar arrangement may be used for the measurement
of nonlinear Thomson photons in a vacuum chamber.

The scheme of Fig.\ref{fig:hrs} is the simplest one  can envision, but it may be possible
to upgrade it in order to increase the efficiency and/or anticipate possible problems. One
possibility is to include a second device like the one in the figure at a different angle in the 
transverse plane. Another option is to include a multi-mode optical fiber in order to couple the
outcome of the optical system to the photon detector. This would allow to place the photon counter away
from the experimental zone in order to shield it from eventual secondary radiation, e.g., X-rays, and to reduce 
undesired background.

Hereafter, we will assume that the optical system can be
parametrized by its {\it field of view} (FoV), namely, the size of the $z$-region from
which scattered photons can be collected, and its {\it numerical aperture} (NA), 
which provides an angular cut for the photons entering the optical system.
We also assume that the {\it depth of field} (DOF), namely the size of the region that is transverse to the laser beam in the direction of the photon-collecting lens system, does not restrict
the detected signal. This approximation is justified if DOF $\gg w_0$.
We stress that it is only necessary to count the number of photons and not to resolve
the location where they were originated.

\subsection{The field of view and geometric efficiency}
\label{sec: FOV}

\begin{figure}[htb]
\includegraphics[width=0.49\textwidth]{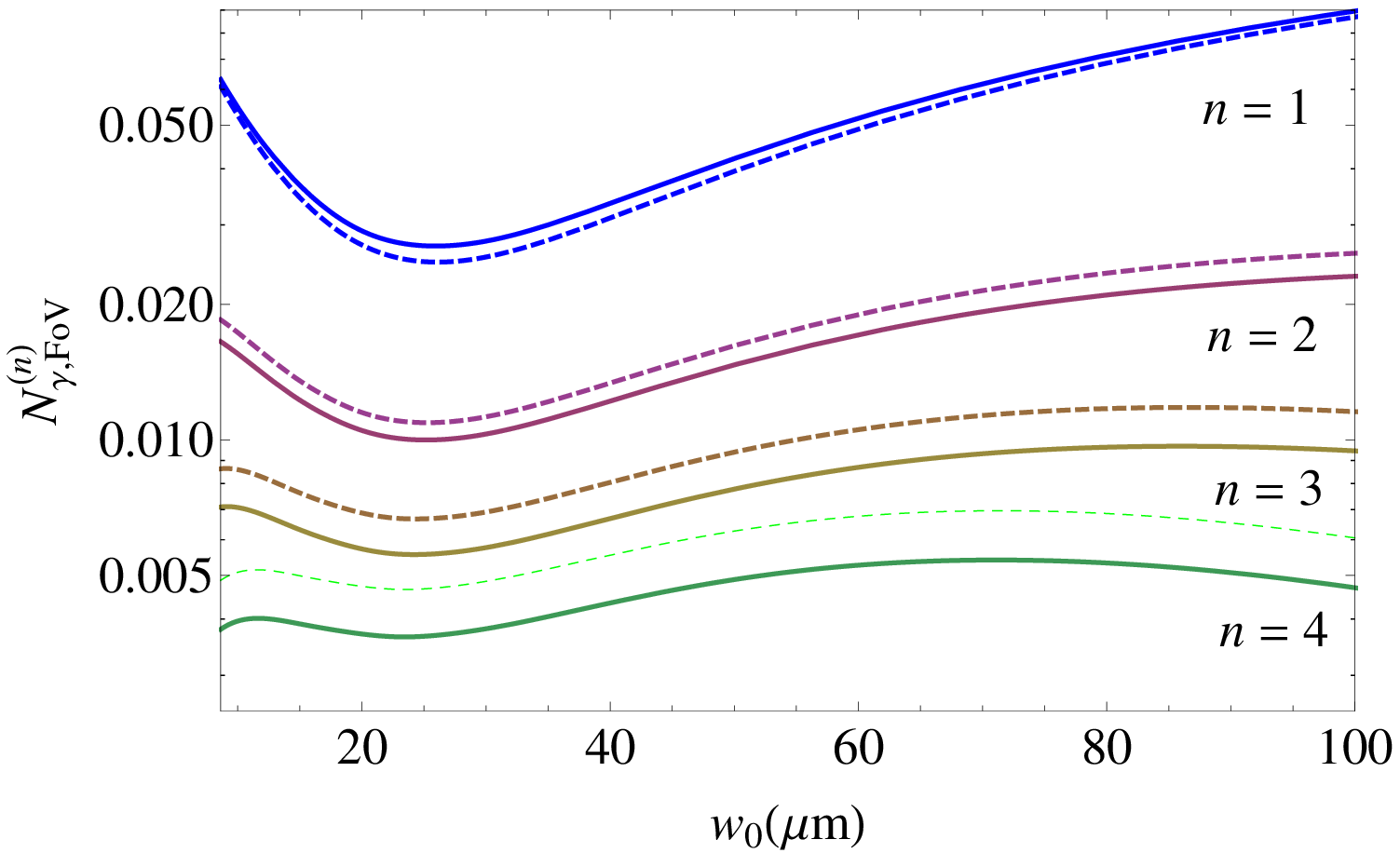}
\includegraphics[width=0.49\textwidth]{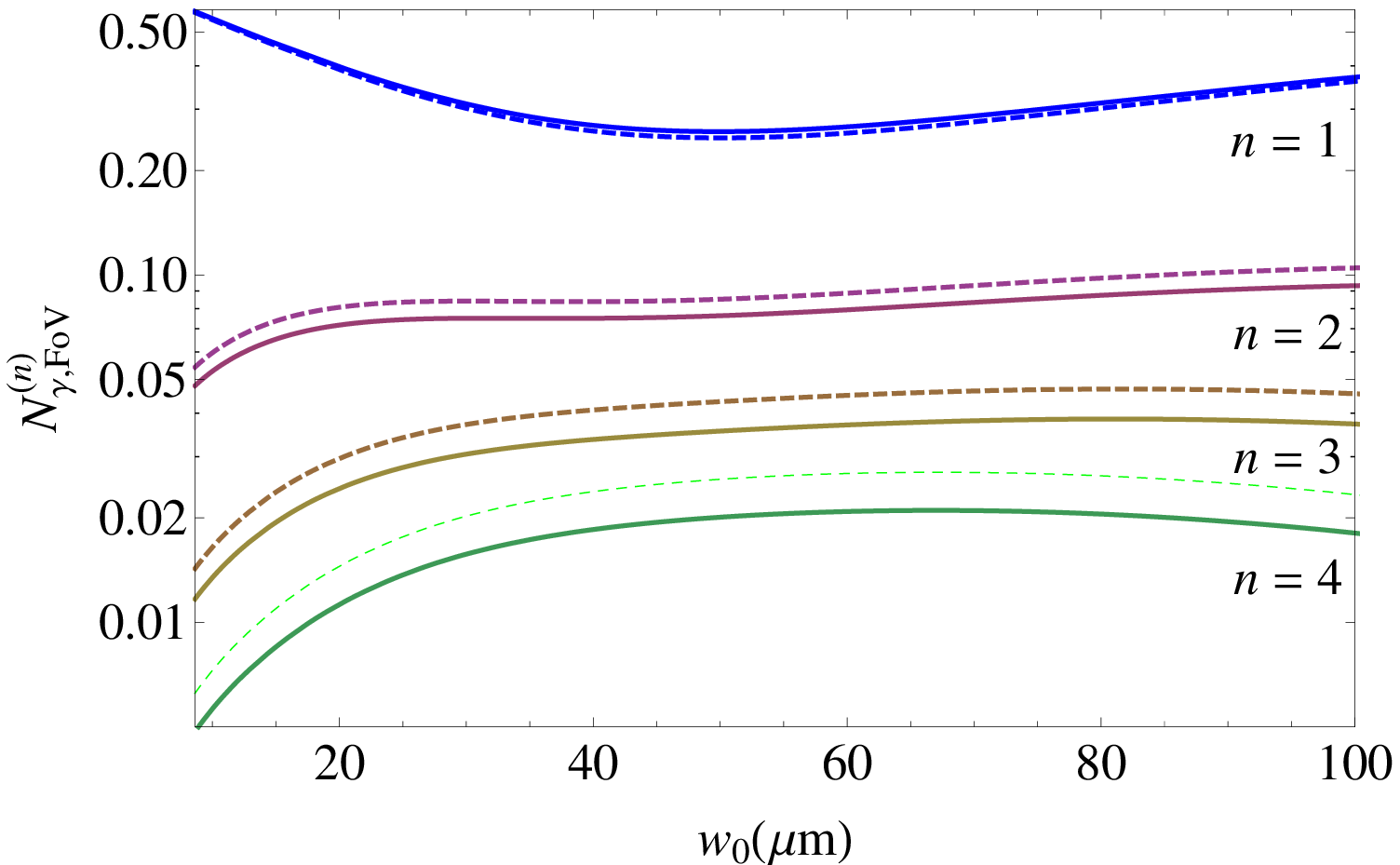}
\caption{ The number of scattered photons 
within the region $z\in (-z_m,z_m)$
from a Gaussian pulse
as a function of the beam waist. The parameters of the laser
beam and $n_e$ are fixed as in Fig. \ref{figNg}. The plot on the left corresponds
to
$z_m=5$mm and the one on the right to $z_m=20$mm. Again, the solid lines correspond
to linear polarization and the dashed lines to circular polarization. The y-axis (x-axis) is given in logarithmic (linear) scale.
}
\label{figNgdr}
\end{figure}

Only a fraction of the photons given in Eq. (\ref{Ngamma})
are scattered inside the FoV of the detection system.
Assuming that the center of the active detection region coincides with
the beam focus, we obtain
\begin{equation}
N_{\gamma,FoV}^{(n)}= {\cal K} \int_{-\xi_m}^{\xi_m}
\int_0^\infty \rho\,{\Gamma}^{(n)}(q)   d\rho d\xi.
\label{Ngammadr}
\end{equation}
The integration in $\rho$ is still formally taken up to infinity since typically
the beam is concentrated in a submillimeter region in the transverse plane which
is assumed to be within the DOF of the collection system in that direction.
Our goal in this section is to estimate the geometric efficiency factor associated
to the finiteness of the FoV $N_{\gamma,FoV}^{(n)}/N_{\gamma}^{(n)}$
 and to discuss the role of $w_0$. 
 Notice that $w_0$ enters the expression in Eq. (\ref{Ngammadr})
in three different ways: in the expression of ${\cal K}$ (recall Eq. (\ref{calK})),
in the value of $q_0$ which affects $q$ through Eq. (\ref{q2}) and in the
value of the limit value of the integral $\xi_m = z_m / z_R = z_m \lambda_0 /
\pi w_0^2$ where $z_m$ is half the FoV. Two examples of the numerically 
computed value of $N_{\gamma,FoV}^{(n)}$ as a function of $w_0$ are
presented in Fig. \ref{figNgdr}.

As compared to Fig. \ref{figNg}, we can observe that the dependence of 
$N_{\gamma,FoV}^{(n)}$ on $w_0$ is much milder than that of $N_{\gamma}^{(n)}$.
Qualitatively, this can be understood as follows for $n>1$: for moderate values of $w_0$,
the integral in 
Eq. (\ref{Ngamma}) is, roughly, proportional to $q_0^3 \propto w_0^{-3}$ (multiplying by
the prefactor ${\cal K}\propto w_0^4$, we find $N_{\gamma}^{(n)} \propto w_0$).
The reason is that the integral is proportional to the volume (in $\rho-\xi$ coordinates) 
of the region
where significant dispersion takes place. The limiting values of $\xi$ and $\rho$ are proportional
to $q_0$ and thus the volume is proportional to $q_0^3$ \cite{pressure}.  When the
integral is cut as in Eq. (\ref{Ngammadr}) and $\xi_m$ lies within the significant dispersion region,
the integral in $\rho$ still picks up a $q_0^2\propto w_0^{-2}$ factor and the integral in $\xi$
is roughly proportional to $\xi_m \propto z_R^{-1} \propto w_0^{-2}$. Thus, the $w_0^4$ 
in ${\cal K}$ cancels out
the $w_0^{-4}$ from the integral and the dependence of $N_{\gamma,FoV}^{(n)}$ on $\omega_0$ is approximately
flat. When $w_0$ is very small, however, $\xi_m$ becomes very large meaning that the region where $q$ is
large enough to have significant harmonic production is very small and is comprised within
$\xi < \xi_m$. Then  $N_{\gamma,FoV}^{(n)} \propto w_0$  for small $w_0$.
This can be appreciated in the plot of the right in Fig. \ref{figNgdr}.

\begin{figure}[htb]
\centerline{\includegraphics[width=0.6\textwidth]{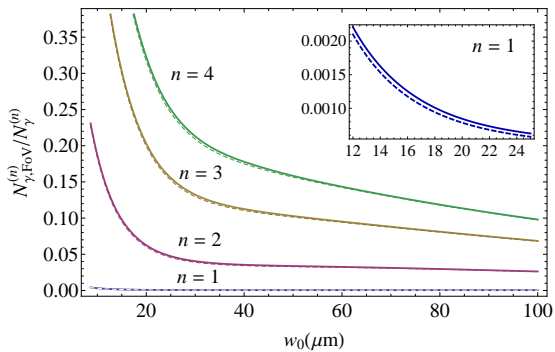}}
\caption{ Fraction of photons scattered within the FoV.
The physical parameter are those of Fig. \ref{figNg} together with $z_m=5 $mm.
The region $w_0< 25\mu$m is enlarged for $n=1$ in the inset.
}
\label{figquotient}
\end{figure}

Finally, it is interesting to plot the fraction of photons which
are indeed scattered within the FoV of the detection system,
namely the geometric efficiency associated to the FoV,
\begin{equation}
\epsilon_{FoV}^{(n)}=\frac{N_{\gamma,FoV}^{(n)}}{N_{\gamma}^{(n)}}.
\end{equation}
Notice that this  is just the quotient of the quantities plotted in 
Fig. \ref{figNgdr} and Fig. \ref{figNg}.  A sample computation is displayed
in Fig. \ref{figquotient}. The fraction  is larger for higher harmonics
because the scattering region gets more and more concentrated around the beam focus.
This plot permits to give an order of magnitude for one of the factors entering the geometric
efficiency. For instance, for $n=2$, $w_0=15\mu$m with $E_{pulse}=30$J, $\tau=30$fs, $\lambda_0=800$nm,
FoV=10mm, a 10\% photons are scattered from the region from which they do enter the light-collection system. Obviously, this depends strongly on the FoV itself, see Fig. \ref{figNgdr}.

\begin{figure}[htb]
\centerline{\includegraphics[width=0.6\textwidth]{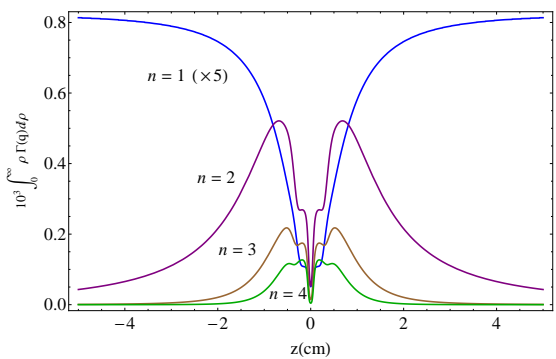}}
\caption{ Distribution along the longitudinal direction $z$ of the scattered
radiation. The plot has been made taking $E_{pulse}=30$J, $\lambda_0=800$nm,
$\tau=30$fs, $w_0=15\mu$m. In order to present all plots in the same graph, the $n=1$ profile
was divided by 5.
}
\label{fig:bump}
\end{figure}

For simplicity, up to now we have assumed the center of the photon collection system to be
at $\xi=0$. However, as a direct consequence of considering the evolution of the Gaussian  beam profile,
this is not always the optimal choice. In order to get some qualitative insight, let us model
(for $n>1$)
$\Gamma^{(n)}(q)\approx b_n \Theta(q-q_{step,n})$ where $b_n, q_{step,n}$ are constants
and $\Theta(x)$ is Heaviside function. Then, it is straightforward to check
that the quantity $\int_0^\infty \rho\,{\Gamma}^{(n)}(q)   d\rho$  has two maxima at
$\xi_c \approx \pm \sqrt{\frac{q_0^2}{e\,q_{step,n}^2-1}}$.  This simple argument 
qualitatively captures the behaviour depicted in Fig. \ref{fig:bump}, which can be found by direct numerical integration.

\begin{figure}[htb]
\includegraphics[width=0.49\textwidth]{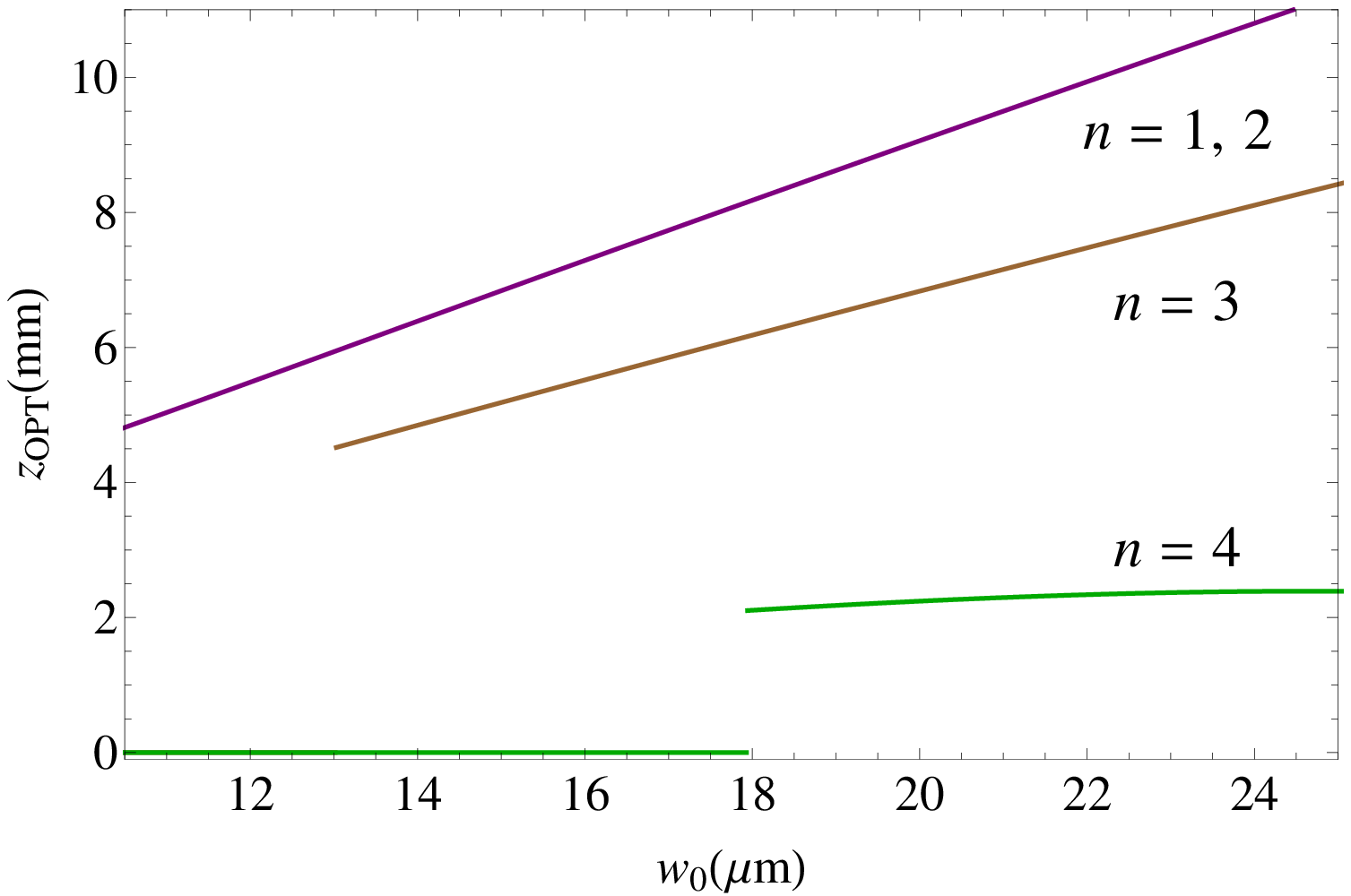}
\includegraphics[width=0.49\textwidth]{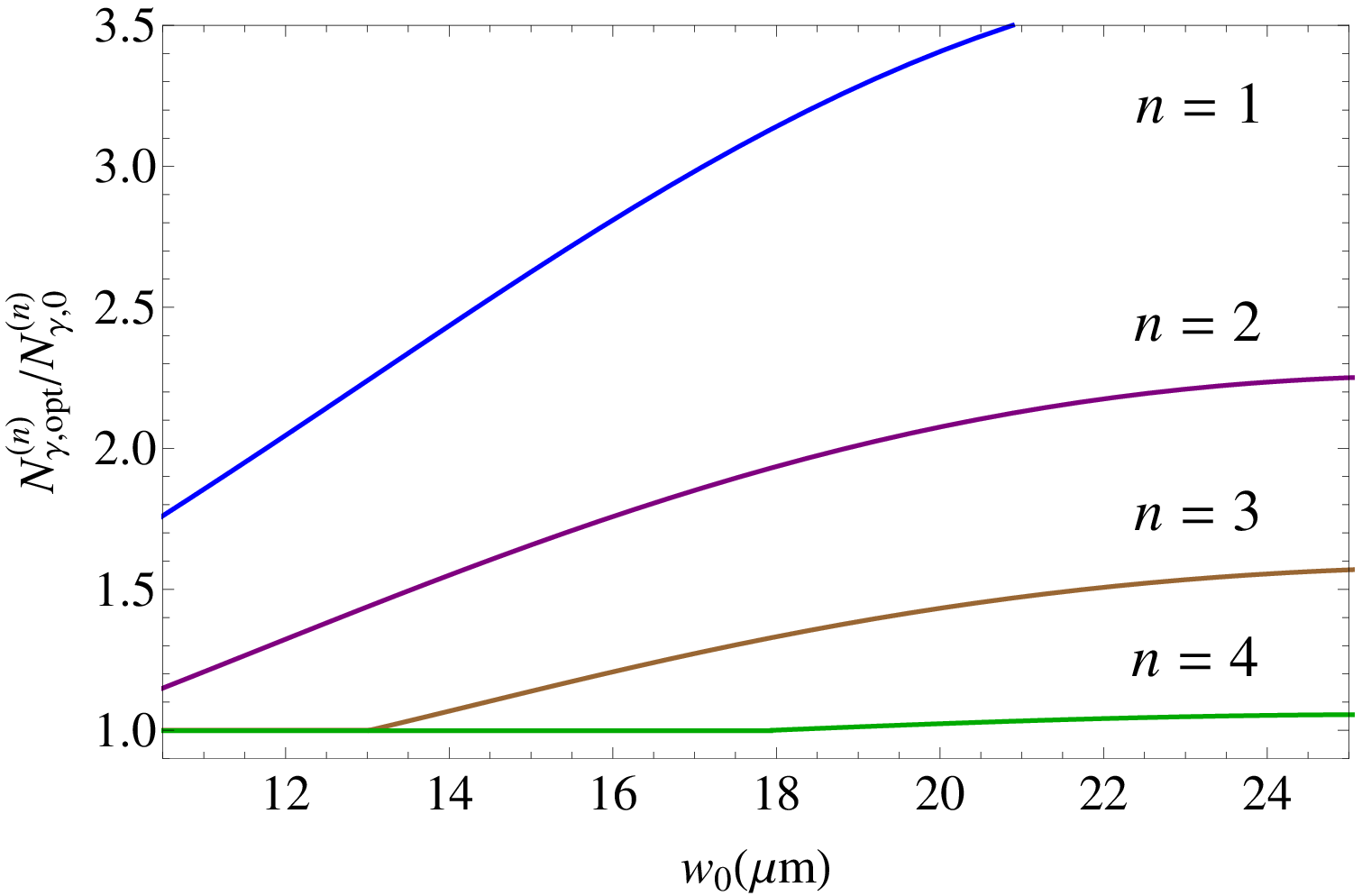}
\caption{ Left: Optimal value of $z_c$ for the different harmonics as a function
of the waist radius. Right: Plot of the factor by which efficiency is increased, as compared to placing
the detector centered around $z=0$. Parameters are as in Figs. \ref{figNg}, \ref{figquotient}.
}
\label{fig:displaced}
\end{figure}

By placing the detection system
around the corresponding $\xi_c$, we obtain 
\begin{equation}
N_{\gamma,FoV}^{(n)}= {\cal K} \int_{\xi_c-\xi_m}^{\xi_c+\xi_m}
\int_0^\infty \rho\,{\Gamma}^{(n)}(q)   d\rho d\xi,
\end{equation}
 which is in general larger than
the expression of Eq. (\ref{Ngammadr}). Thus, by appropriately displacing the detection system
with respect to the beam focus, the geometric efficiency factor can be increased to some extent.
We have performed an analysis of an example, using the actual values of $\Gamma^{(n)}(q)$. The results showing
the optimal value of $z_c = \xi_c \pi w_0^2 / \lambda_0$ and the increase in the geometric efficiency
as compared to Fig. \ref{figNgdr} are displayed in Fig. \ref{fig:displaced}.

Formally, for $n=1$, the optimal value of $z$ would be $z_c\rightarrow\infty$, since the associated profile
asymptotes to a constant in Fig. \ref{fig:bump}. However, for convenience, $z_c$ corresponding to $n=1$ will be taken to coincide with that for $n=2$.
On the other hand, the discontinuities exhibited by the curves $n=3,4$ in Fig. \ref{fig:displaced} can be explained by the following argument. 
The displacement of the detector is useful to capture one of the bumps of the radiation distribution, as
shown in Fig. \ref{fig:bump}. However, when the bumps for both positive and negative values of $z$ come close enough
and, as a consequence, can be included within the FoV of the detector, the optimal choice is simply
to take $z_c=0$. In particular, for a FoV$=10$mm, that would be the case of the $n=4$ lines of Fig. \ref{fig:bump}.

The optimized geometric efficiency associated to the FoV is then given by the result
of Fig. \ref{figquotient} multiplied by the enhancement factor that can be achieved by displacing
the detector with respect to the focus (see Fig. \ref{fig:displaced}). For instance, let us consider a laser
 featuring $E_{pulse}=30$J, $\tau=30$fs, $\lambda=800$nm  $w_0=15\mu$m and a detector with
FoV=10mm. For such a system, the efficiency for the collection of $n=2$ photons would then be optimum (e.g., $\epsilon_{FoV}^{(2)}\approx 0.16$ in this particular case)
for $z_c=6.85$mm.

\subsection{Numerical aperture. Angular acceptance}
\label{sec: apert}

Among all photons coming from the FoV, only those within certain
angular cuts are effectively captured by the collection system.
The quantity defining the  angular acceptance of the optical system is NA. Assuming 
that the refractive index of the medium is 1, it is defined
as NA $= \sin \tilde \theta_i$, such that the radiation with $\tilde \theta < \tilde
\theta_i$ is measured. $\tilde \theta$
is defined as the angle between the photon direction and the axis joining the
scattering region to the center of the optical system and thus is not the
$\theta$ used in the previous sections. 
The goal of this section is to estimate the fraction of scattered
photons lying within the NA, thus finding the corresponding factor for the
geometric efficiency of the photon collection system.

The number of photons scattered per unit solid angle by a Gaussian beam traversing a 
vacuum chamber is easily derived from the expressions given in section \ref{sec II},
\begin{equation}
\frac{dN_\gamma^{(n)}}{d\Omega}={\cal K}\int_{-\infty}^\infty \int_0^\infty 
\frac{1}{n}\rho {\cal M} f^{(n)} d\rho d\xi.
\label{angular}
\end{equation}

The efficiency associated to the numerical aperture would then be
\begin{equation}
\epsilon_{NA}^{(n)}=\frac{2\int_{\tilde \theta<\tilde \theta_i}\left(dN_\gamma^{(n)}/d\Omega\right)d\Omega}
{\int_\Omega\left(dN_\gamma^{(n)}/d\Omega\right)d\Omega},
\label{ENA1}
\end{equation}
where the integral in the denominator is taken over the full solid angle. The factor of 2 in
the numerator comes from considering both the integrals in $\tilde \theta \in [0,\tilde\theta_i]$
and in $\tilde \theta \in [\pi-\tilde\theta_i,\pi]$ because of the mirror placed opposite to the detector
(see Fig. \ref{fig:hrs}).

The expression Eq. (\ref{ENA1}) can be directly evaluated numerically in any particular case. 
In fact, we have verified that the angular distribution of Eq. (\ref{angular})
is rather accurately approximated by the low $q$ angular dependence of the integrand. We can then obtain a simple estimate of $\epsilon_{NA}^{(n)}$, depending only on the
polarization state, the harmonic number, and the numerical aperture. 
Defining
\begin{equation}
\Xi^{(n)}(\theta,\varphi) \equiv \frac{1}{n}\lim_{q\to 0} \frac{f^{(n)}}{q^{2n}},
\label{Xidef}
\end{equation} 
Eq. (\ref{ENA1}) reads
\begin{equation}
\epsilon_{NA}^{(n)}\approx\frac{2\int_{\tilde \theta<\tilde \theta_i}\Xi^{(n)}d\Omega}
{\int_\Omega\Xi^{(n)}d\Omega}.
\label{ENA2}
\end{equation}
The cases of circular and linear polarization are discussed separately below.

\subsubsection{Circular polarization}

By expanding Eq. (\ref{fncirc}), we find
\begin{equation}
\Xi_c^{(n)}=\frac{2^{2-3n}n^{2n-1}}{(n-1)!}(1+\cos^2\theta)(\sin \theta)^{2n-2},
\label{Xicirc}
\end{equation}
which gives
\begin{equation}
\int_\Omega\Xi_c^{(n)}d\Omega = \frac{2^{4-n}\pi(1+n)n^{2n}}{(2n+1)!}.
\end{equation}
In order to compute the numerator of Eq. (\ref{ENA2}), let us define a new set of spherical coordinates
obtained by a rotation of $\pi/2$ with respect to the $x$-axis,
\begin{eqnarray}
\cos \theta &=& \sin\tilde \theta \sin \tilde \varphi\,\,,\nonumber\\
\tan \varphi &=& - \cot \tilde \theta \sec \tilde \varphi \,.
\label{coord1} 
\end{eqnarray}
This amounts to placing the detection system ($\tilde \theta=0$) along the $y$-axis.
By inserting the expression given in Eq. (\ref{coord1}) into Eq. (\ref{Xicirc}) and Eq. (\ref{ENA2}), estimates
for $\epsilon_{NA}^{(n)}$ can be found. For instance, if $NA=0.5$, meaning $\tilde\theta_i = \pi/6$, we obtain
$\epsilon_{NA}^{(1)}\approx 0.11$, $\epsilon_{NA}^{(2)}\approx 0.17$, 
$\epsilon_{NA}^{(3)}\approx 0.20$, $\epsilon_{NA}^{(4)}\approx 0.23$.

\subsubsection{Linear polarization}

This case is more complicated than the previous one because of the $\varphi$-dependence
of the differential cross section and the cumbersome form of Eq. (\ref{fn}).
The angle $\beta$ between the polarization direction and the location of the detector
has to be properly chosen.
All computational details are relegated to appendix B, whereas only
the estimates of the geometric efficiency factor related to angular acceptance  are quoted here. 
Assuming NA=0.5, namely  $\tilde\theta_i = \pi/6$,
they are 
$\epsilon_{NA}^{(1)}\approx 0.19$, $\epsilon_{NA}^{(2)}\approx 0.19$,  
$\epsilon_{NA}^{(3)}\approx 0.31$ and $\epsilon_{NA}^{(4)}\approx 0.49$.
Notice that they are larger than the efficiencies that can be achieved with circular
polarization. This is due to the breaking of the azimuthal symmetry, implying that the distribution of
scattered power is more inhomogeneous over the solid angle. We can profit from this fact
by suitably choosing the location of the photon collection system.

\subsection{Summary and an example}

Let us summarize the main results of section \ref{sec: collect}. Once given the characteristics of a
laser pulse ($E_{pulse}$, $\tau$, $\lambda_0$) and of a photon collection system
(its FoV and NA),  the waist radius of the
beam and the position of the detector, both in the longitudinal direction $z_c$ and its
angular position in the transverse plane $\beta$ can be optimized.
In section \ref{sec: apert} and appendix B, we give an estimate of the optimal $\beta$ and the efficiency associated to
the angular acceptance.
 Section \ref{sec: FOV} discusses how to choose
values of $w_0$ and $z_c$ and gives the quantitative results for the geometric 
efficiency associated to the FoV in a sample case. It does not seem possible to
provide a simple estimate for these quantities as a function of all the input parameters.

It is worth mentioning that the full geometric efficiency is not exactly the product of
$\epsilon_{FoV}$ and $\epsilon_{NA}$ since, in an actual computation, both cuts should be
taken into account simultaneously. However, we have discussed their computations separately
for clarity of exposition. In any case, the error we make by splitting the computation
in this fashion is not large, although it depends on the particular case. 
For instance, for the quoted case with linear polarization with
$E_{pulse}=30$J, $\tau=30$fs, $\lambda=800$nm,
FoV = 10mm, $w_0=15\mu$m, $z_c=6.85$mm, $NA=0.5$, $\beta=0.84$ the efficiencies given above
are 
$\epsilon_{FoV}^{(2)}=0.163$ and $\epsilon_{NA}^{(2)}=0.19$, such that $\epsilon_{FoV}^{(2)}
\epsilon_{NA}^{(2)}=0.031$.
This should be compared with the computation including directly in the integrals both cuts
which gives $\epsilon_{geom}=0.030$.

\section{Some quantitative estimates}
\label{sec: quant}

One of the conclusions of the previous sections is that the beam polarization 
affects only mildly the number of scattered photons, see Figs. \ref{figNg}-\ref{fig:displaced}.
In contrast, it modifies more severely the angular distribution of radiation and, therefore, the
number of photons that propagate within the numerical aperture of the detector. 
This distribution is more inhomogeneous for linear polarization and this fact 
permits to enhance the geometric efficiency by suitably placing the detection system,
see section \ref{sec: apert}. In the following we will concentrate on linear polarization.
It was shown in \cite{pressure} that in this case 
the number of scattered photons per pulse is
$N_\gamma^{(n)} \approx c_n {\cal K}\,q_0^3$
 where the $c_n$ are coefficients that can be computed numerically,
 $c_1\approx 275$, $c_2\approx 1.3$, $c_3\approx 0.22$, $c_4\approx 0.088$.
 This expression is valid for large $q_0$, corresponding to the region of small $w_0$
  in Fig. \ref{figNg}, in which
 $N_\gamma^{(n)} \propto w_0$. A laser with repetition rate $r_r$ operating
 for a time interval $\Delta t$ produces $\Delta t\,r_r
 $ pulses. 
 Under conditions of XHV, the number of
 detected photons is proportional to the electron density:
\begin{equation}
N_{\gamma,det}^{(n)} = {\cal A}\,n_e.
\label{Ngresult2}
\end{equation}
The value of the proportionality constant can be found by
 substituting the values for ${\cal K}$ and $q_0$ in the expression for $N_\gamma^{(n)}$ given above \cite{pressure}, so that
\begin{equation}
{\cal A} \approx \frac{4c_n}{\pi} (\Delta t\,r_r)
 \alpha\,\frac{w_0 \lambda_0 r_0^{3/2}}{(c\,\tau)^{\frac12}}
\left(\frac{E_{pulse}}{m_e c^2}\right)^{\frac32}f\,\kappa\,.
\label{Ngresult3}
\end{equation}
The parameter $\kappa$ is the correction due to a non-trivial time envelope
and will be fixed to a typical value 0.8, see section \ref{sec: envelope}. The efficiency factor
 $f \approx \epsilon_{geom} \epsilon_q \epsilon_\lambda$ is the efficiency factor 
 including the geometric efficiency (see section \ref{sec: collect}), the quantum efficiency of the detector
 and the cuts imposed by the frequency filter.
 
 Furthermore, $n_e$ is proportional to the pressure,
 \begin{equation}
 n_e = \eta\frac{p} {k_B T},
 \label{eta}
 \end{equation}
where $\eta$ is the 
average number of weakly bound electrons per molecule \cite{pressure}
--- namely, those in the barrier suppression
regime. It depends on the atomic and molecular composition of the remnant gas in the vacuum chamber.
In a canonical XHV, its value would be $\eta \approx 2$ since it is mostly composed of
hydrogen molecules \cite{bryant1965,fernandez2012}.

The goal of this section is to provide estimates of these quantities for three PW facilities that will be available in the near future, namely VEGA \cite{VEGA}, JuSPARC \cite{JuSPARC,JuSPARC2}, and a 10 PW branch of the ELI project \cite{ELI}, which have been chosen because of
their sizable repetition rates, see table \ref{tab:1}.  
\begin{table}[h]  
\footnotesize
\begin{center}
\begin{tabular}{|c|c|c|c|c|c|}
\hline
\textbf{Facility} & \textbf{$P_p$} (PW) & 
\textbf{$E_{pulse}$} (J) &
\textbf{$\tau$} (fs) 
& \textbf{$\lambda_0$} (nm) & \textbf{$r_r$} (Hz)
\\
\hline
VEGA & 1 & 30 & 30 & 800  & 1
  \\
\hline
JuSPARC & 1.5 & 45 & 30 & 800  & 1
  \\
\hline
ELI 10 PW & 10 & 300 & 30 & 800 & 0.1
  \\
\hline
\end{tabular}
\caption{A few facilities that will operate in the near future.}
\label{tab:1}     
%$^a$ Table foot note (with superscript)
\end{center}
\end{table}

In all cases, we will assume band-pass filters for each harmonic as $\lambda_1$(nm) $\in [800,1200]$,
$\lambda_2$(nm) $\in [400,800]$, $\lambda_3$(nm) $\in [267,400]$, $\lambda_4$(nm) $\in [200,267]$
(recall that the photon wavelength is shifted by a $q$-dependent factor, Eq. (\ref{lambdan})). This
choice should be adjusted for a particular detector and frequency filter, see an enlarged discussion in section
 \ref{sec: discussion}.
We consider 
an experiment running for $\Delta t=$ 1 day. 
A detection system with FoV = 10mm, NA = 0.5 and an average quantum
efficiency of $\epsilon_q=0.25$  within the allowed wavelength bands will be considered.
These are sample values intended to be representative and to provide a reasonable
estimate for realistic situations.

The results are summarized in table \ref{tab:2}. Harmonics $n=1,\dots,4$ are considered in
each case, the position of the detector along $z$ is optimized as explained in section
\ref{sec: FOV} and the angle $\beta$ is chosen in each case as in appendix B. The waist radii,
 chosen to comply with (\ref{w0cond}), are taken to be
 $w_0=15\mu$m, $w_0=15\mu$m, $w_0=27.5\mu$m for VEGA, JuSPARC and ELI 10, respectively,
($q_0\approx 11.5$, $q_0\approx 14.1$, $q_0\approx 19.8$). 
The efficiency factors $\epsilon_{geom}$, $\epsilon_\lambda$ and the proportionality factor ${\cal A}$
of equations (\ref{Ngresult2}), (\ref{Ngresult3}) are found by computing the appropriate numerical
integrals. 

\begin{table}[h]  % Give a unique label
\footnotesize
\begin{center}
\begin{tabular}{|c|c|c|c|c|c|}
\hline
\textbf{Facility} & n & \textbf{$z_c$} (mm) & 
\textbf{$\epsilon_{geom}$}  &
\textbf{$\epsilon_\lambda$}
& \textbf{${\cal A}$ }(mm$^3$) 
\\
\hline
VEGA & 1 & 6.85 & $6.5\times 10^{-4}$  & 0.94  & 113
  \\
\cline{2-6}
 & 2 & 6.85 &  $3.0\times 10^{-2}$ & 0.96  & 25
  \\
\cline{2-6}
 & 3 & 5.2 &  $7.2\times 10^{-2} $& 0.68  & 7.6
  \\
\cline{2-6}
 & 4 & 0  &  $9.6\times 10^{-2}$ & 0.16  & 0.9
  \\
\hline
JuSPARC & 1 & 8.4 & $ 5.4\times 10^{-4}$ & 0.94  & 172 
  \\
\cline{2-6}
 & 2 & 8.4 & $2.6\times 10^{-2} $ &  0.97 & 41 
  \\
\cline{2-6}
 & 3 & 6.4 &  $6.8\times 10^{-2} $& 0.65  & 12 
  \\
\cline{2-6}
 & 4 & 2.35 &  $7.2\times 10^{-2}$  & 0.19  & 1.5 
  \\
\hline
ELI 10 & 1 & 4.35 &  $1.2\times 10^{-4}$ & 0.95  & 122 
  \\
\cline{2-6}
 & 2 & 4.35 &  $6.2\times 10^{-3}$ & 1.0  & 32
  \\
\cline{2-6}
 & 3 & 3.3 &  $1.8\times 10^{-2}$ & 0.53  & 8.4 
  \\
\cline{2-6}
 & 4 & 1.25  &  $1.1\times 10^{-2} $ & 0.12  & 0.46
  \\
\hline
\end{tabular}
\caption{Estimates for three future facilities.}
\label{tab:2}     

\end{center}
\end{table}

The first observation is that, even if the geometric efficiency is much lower
for $n=1$, the majority of photons reaching the detector are of this fundamental
harmonic. Nevertheless, the difference is less than one order of magnitude with respect
to $n=2$. Gauging the pressure by looking at this second harmonic would have several assets:
it would help to avoid possible undesired background of photons from the main beam reaching the detector
without having been 'Thomson scattered' and also to reduce other sources of background such as
thermal noise. Moreover, photon detectors typically reach higher quantum efficiencies with smaller
dark counts in the visible than in the IR, although  
that can depend on the detector itself, see \cite{eisaman} for a review of single photon detectors. 
Recall that in table \ref{tab:2}, the same quantum efficiency was assumed in 
all cases. On the other hand, the separate measurement of {\it both} the $n=1$ and $n=2$ harmonics can be used to self-calibrate the procedure.

Let us first estimate the minimum pressure that could be gauged, in principle,
in a one day experiment at the three mentioned facilities by detecting the $n=1$ photons. Since extreme vacuum is mostly formed by molecular
hydrogen, we take $\eta=2$ in Eq. (\ref{eta}). We require that the average number of photons measured
in the detection period is at least 10. Then $p_{min}\approx 10 k_B T / (2 {\cal A})$, where
the values of ${\cal A}$ are given in table \ref{tab:2}. For VEGA, we obtain
$p_{min}\approx 1.8\times 10^{-13}$Pa at room temperature $T=300$K or 
$p_{min}\approx 2.3\times 10^{-15}$Pa at liquid He temperature $T=4$K.
For JuSPARC, 
$p_{min}\approx 1.2\times 10^{-13}$Pa at $T=300$K or 
$p_{min}\approx 1.5\times 10^{-15}$Pa at $T=4$K.
For ELI 10, 
$p_{min}\approx 1.6\times 10^{-13}$Pa at $T=300$K or 
$p_{min}\approx 2.2\times 10^{-15}$Pa at $T=4$K.
It should be noted that these results may be improved, leading to the possible measurement of even lower pressures, by using a different setup allowing for a greater geometric efficiency. The theoretical limit can be found by multiplying the results of Ref. \cite{pressure} including the time envelop correction that we have computed above. In any case, the optimization procedure that we have developed above can be straightforwardly generalized to any given geometry.

Let us consider the case in which the second harmonic, $n=2$, is used to gauge the vacuum, and find the estimates of the minimum pressure that could be gauged
in one day in the three mentioned facilities. For VEGA, we obtain
$p_{min}\approx 8\times 10^{-13}$Pa at room temperature $T=300$K or 
$p_{min}\approx 1.1\times 10^{-14}$Pa at liquid He temperature $T=4$K.
For JuSPARC, 
$p_{min}\approx 5\times 10^{-13}$Pa at $T=300$K or 
$p_{min}\approx 0.7\times 10^{-14}$Pa at $T=4$K.
For ELI 10, 
$p_{min}\approx 6.5\times 10^{-13}$Pa at $T=300$K or 
$p_{min}\approx 0.9\times 10^{-14}$Pa at $T=4$K.

\section{Discussion}
\label{sec: discussion}

In this section, a few interesting questions that have been left out of
the general discussion are addressed. 

For linear polarization, the trajectory
of an electron extracted from an atom by the electromagnetic field passes near the ion during
its oscillation, opening the possibility of electron-nucleus recombination with the associated
photon emission. This harmonic-generating 
phenomenon has not been taken into account in the discussion. The safest
possibility is to introduce a slight ellipticity in the beam polarization in order to reduce the probability of
this circumstance to happen,
 while keeping nearly unchanged the angular distribution of the Thomson radiation.

We have discussed the minimum pressure that can be gauged in a given situation, associated to having a 
detectable signal of photons. Another interesting question is which would be the {\it maximum}
measurable pressure. The method presented in this note would be useful as long as the pressure and the
number of scattered photons remain proportional to each other. This can only break down when the 
density of active electrons is high enough to introduce collective effects. A extremely conservative
estimate
would be to compare the volume per electron ($n_e^{-1}=k_B T/\eta p$) to the volume 
of the 
laser pulse (roughly $\frac{\pi}{2}w_0^2 c\,\tau$). Taking values
$w_0=15\mu$m, $\tau=30$fs, $T=300$K, $\eta=2$ gives $p\approx 10^{-6}$Pa. 
This value is in the so-called high vacuum regime in which pressure can be measured with
great precision with standard techniques. In fact, comparing in this regime laser measurements with
standard ones would be a valuable benchmark calibration of the method. 

It is conceivable to design ultra-high or extreme vacuum gauges using table-top terawatt lasers rather
than PW facilities. These could find more applications since the cost of the required device would
be orders of magnitude lower.
The reduced power would be compensated, at least partially, by larger repetition rates. However, even if
the general idea presented here would hold, the actual computations would not. For beams far from 
the diffraction limit, terawatt lasers 
would yield $q_0<1$, i.e., intensities out of the relativistic regime. For instance, with
$P_p=3$TW, $w_0=15\mu$m, $\lambda_0=800$nm, we obtain $q_0^2\approx 0.4$. Harmonic production
would be suppressed and expressions like
Eqs. (\ref{Ngresult2})-(\ref{Ngresult3}) would fail. Moreover, for pulse durations down to the few-cycle limit, the approximation of slowly-varying envelope considered throughout this paper would no longer hold --- see
for instance \cite{mackenroth}. 
The exploration of such limiting case, although interesting, 
lies beyond the scope of the present work.

Finally, it is worth discussing the wavelength spectrum of the scattered photons. 
In laboratory frame, the spectral distribution that can be computed with the 
expression used above is rather broad \cite{pressure}. It would be further broadened
by at least two additional effects which are enhanced for short pulses: the width of the 
incoming laser pulse itself and the departure from the results of \cite{sarachik} when
the envelope is not slowly-varying \cite{krafft,Gao}. 
The spectra for the different harmonics can be overlapping, producing a sort of supercontinuum.
In fact, splitting the results in harmonics is just a convenient computational artifact, while
the physical measurable result is the sum of all them. In that sense, the results presented in
table \ref{tab:2} are lower limits since they only include the first (larger) contribution in 
the harmonic sum for the different wavelength bands. Notice that this overlap is harmless for the
proposed pressure gauge since the total signal remains proportional to the number of scattering
electrons.
It is obvious that photon detectors with broad
efficiency curves would be necessary. Once given a curve in a particular case, 
the computations shown above
can be generalized by properly including it in the integrals, instead of assuming a constant
quantum efficiency and a sharp band-pass filter.

\section{Conclusions}
\label{sec: conclusions}

The availability of ultra-short and ultra-intense laser pulses opens the possibility of gauging extreme vacuum pressure by photon counting. The huge photon concentration in these pulses allows to overcome, in the long run, the scantiness of scattering centers in extremely rarefied gases. The shortness of the pulses allows to synchronize the measurements with the pulse passage and to eliminate (or, at least, dramatically reduce) the undesired background 
by gating in time the signal produced by the photon detectors. Moreover, for the high intensities that can be obtained with focused PW laser beams corresponding to $q \gtrsim 1$, a significant quantity of radiation is non-linearly Thomson-scattered in harmonics $n>1$. The selective detection of only these higher harmonics can be used to significantly reduce any possible background coming from the possible deviations of the original beam from the axially-centered Gaussian distribution.
We have considered a typical photon
collection system and shown how to optimize the vacuum gauge accuracy by properly placing the detectors. Within this realistic setup, we have obtained optimized geometric efficiencies of the order of a few percent
for $n>1$. 
We have also shown that these results hold for any choice of polarization of the incoming pulse, with numerical variations of the order of the unity. With these assumptions,  pressures of the order of $p=10^{-13}-10^{-12}$ Pa at room temperature can be
measured in a one-day experiment at VEGA, JuSPARC or ELI 10, 
assuming that such conditions can be created and maintained during this time. This same procedure can be also applied to more encompassing dispositions of the detectors, that can lead to greater geometrical efficiencies and may eventually allow to lower the limiting pressure that can be achieved.

Upgrading and understanding the classical
vacuum may be crucial for experiments trying to explore properties of the quantum vacuum \cite{qvac-reviews,PPSVsearch,new_physics}. Apart from gauging the pressure, nonlinear Thomson scattering might also be useful for beam 
characterization \cite{krafft} since its detection can be a probe of the focusing region where it
is impossible to introduce any direct characterization system. 
Har-Shemesh and Di Piazza have proposed to employ it
to provide indirect measurements of the peak intensity \cite{peak} and, in the same spirit, the possibility of studying
beam profiles or time envelopes is worth investigating. Hopefully, the computations
presented here could be instrumental in this direction.

%-------------------------------------------------------------------------
\subsection*{Acknowledgements}
We thank M. B\"uscher, D. Gonz\'alez-D\'\i az, J. Hern\'andez-Toro, J. A. P\'erez-Hern\'andez, 
A. Peralta, L. Roso and C. Ruiz for useful discussions. A.P. is supported by the Ram\'on y Cajal program. 
D.T. thanks the InterTech group of Valencia Politechnical University for hospitality during a research visit that was supported by the Salvador de Madariaga program of the Spanish Government.
The work of A.P. and D.T. is supported by Xunta de Galicia through grant EM2013/002.

%--------------------------------------------------------------------------

\appendix
%%%%%%%%%%%%%%%%%%%%%%%%%%%%%%%%%%%%%%%%%%%%%%%%%%%%%%%%%%%%%%%%%%%%%%%%%%%%%%%%

\setcounter{equation}{0}
\renewcommand{\theequation}{\Alph{section}.\arabic{equation}}
%%%%%%%%%%%%%%%%%%%%%%%%%%%%%%%%%%%%%%%%%%%

\section{Approximate expressions for the $\Gamma^{(n)}(q)$}

The functions $\Gamma^{(n)}(q)$ defined in section \ref{sec: Gamma} are important tools in 
the computation of the number of photons scattered by the nonlinear Thomson 
effect. The formal expressions are rather involved and can only be evaluated
numerically. Nevertheless, we have checked that they can be well approximated
by simple quotients of polynomials
 for the values of $n$  considered, see Fig. 
 \ref{figGamma}. The error of the approximation is under 1\% in most of the range
 and, in fact, plotting the expressions below in Fig. 
  \ref{figGamma} would display lines not distinguishable from the numerical results.
 Only even powers of $q$ are considered since, formally,
  $\Gamma^{(n)}(q)=\Gamma^{(n)}(-q)$. 
  
For linear polarization
\begin{eqnarray}
\Gamma^{(1)}_l(q)&\approx& \frac{8\pi}{3}\frac{q^2(1+0.414 q^2)}{1+1.33 q^2 + 0.497q^4}\,\,,
\nonumber\\
\Gamma^{(2)}_l(q)&\approx& \frac{7\pi}{5}\frac{q^4(1+0.454 q^2)}{1+2.18 q^2 +
1.63 q^4 + 0.539 q^6}\,\,,
\nonumber\\
\Gamma^{(3)}_l(q)&\approx& \frac{207\pi}{224}\frac{q^6}{1+2.97q^2+1.66 q^4 +1.13 q^6}\,\,,
\nonumber\\
\Gamma^{(4)}_l(q)&\approx & \frac{1081\pi}{1620}\frac{q^8}{1+2.79q^2+ 4.79q^4 + 2.07q^6 + 1.05q^8}\,\,.
\nonumber
\end{eqnarray}
For circular polarization
\begin{eqnarray}
\Gamma^{(1)}_c(q)&\approx& \frac{8\pi}{3}\frac{q^2(1+0.249 q^2)}{1+1.20 q^2 + 0.370q^4}\,\,,
\nonumber\\
\Gamma^{(2)}_c(q)&\approx& \frac{8\pi}{5}\frac{q^4(1+0.246 q^2)}{1+1.98 q^2 +
1.34 q^4 + 0.330 q^6}\,\,,
\nonumber\\
\Gamma^{(3)}_c(q)&\approx& \frac{81\pi}{70}\frac{q^6(1+0.245 q^2)}{1+2.74q^2+2.94 q^4 +1.44 q^6
+0.305 q^8}\,\,,
\nonumber\\
\Gamma^{(4)}_c(q)&\approx & \frac{512\pi}{567}\frac{q^8}{1+1.17 q^2+ 7.05 q^4 + 1.55 q^6 + 1.16 q^8}\,\,.
\nonumber
\end{eqnarray}
We have taken into account that the leading term of all $\Gamma^{(n)}(q)$
for small $q$ is of order $q^{2n}$. Its coefficient can be straightforwardly
computed by Taylor expansion and has been inserted in the expressions above.
The rest of coefficients have been fitted to the data found from numerical integrals.

\section{Geometric efficiency related to numerical aperture for 
linear polarization}

It is possible to write down explicit expressions for the
$\Xi^{(n)}$ defined in Eq. (\ref{Xidef}) by expanding (\ref{fn}):
\begin{eqnarray}
\Xi_l^{(1)}&=&\sin^2\alpha\,\,,\nonumber\\
\Xi_l^{(2)}&=&\cos^2\alpha(2\sin^2\alpha- \cos\theta) + \frac18 \sin^2\theta
\,\,,\nonumber\\
\Xi_l^{(3)}&=&\frac{27}{64}\Big(\frac14\sin^2\alpha(1-\cos\theta-6\cos^2\alpha)^2+\nonumber\\
&+& \sin^2\theta \cos^2\alpha + \cos \theta \cos^2\alpha (1-\cos\theta-6\cos^2\alpha)
\Big)\,\,,\nonumber\\
\Xi_l^{(4)}&=&\sin^2\alpha \cos^2\alpha (1-\cos\theta-\frac83\cos^2\alpha)^2+\nonumber\\
&-&\frac14 \cos\theta \cos^2\alpha (1-\cos\theta-\frac83\cos^2\alpha)\nonumber \\
&&(1-\cos\theta-8\cos^2\alpha)+ \nonumber\\
&+& \frac{1}{64}\sin^2 \theta  (1-\cos\theta-8\cos^2\alpha)^2.
\end{eqnarray}
Equivalent expressions for $n=1,2,3$ were given in \cite{sarachik} with two typos
that were corrected in \cite{castillo}.

Since in the case of linear polarization the azimuthal symmetry is broken, it is convenient to choose
the angular position of the detection system with respect to the polarization direction in
order to optimize the detection of photons. Consider a new set of spherical coordinates by
first performing a $\pi/2$-rotation around the $x$-axis followed by a $\beta$-rotation around
the new $y$-axis,
\begin{eqnarray}
\theta&=&\arccos \left(\sin\tilde \theta \,\sin \tilde \varphi\right)\,\,,\nonumber\\
\varphi &=&- \arctan \left(\frac{\sin\beta \sin\tilde \theta\cos\tilde\varphi+\cos\beta
\cos\tilde \theta}{\cos\beta \sin\tilde \theta\cos\tilde\varphi-\sin\beta
\cos\tilde \theta}\right).
\end{eqnarray}
For $\beta=0$, Eq. (\ref{coord1}) is recovered. For a given value of $\beta$ and the
numerical aperture $\sin \tilde \theta_i$, the efficiency (\ref{ENA2}) can be computed.
Then, it is straightforward to find the value of $\beta$ which optimizes $\epsilon_{NA}^{(n)}$
and which indicates where the optical system should be placed. The optimal $\beta$ depends
on the harmonic number and also on the numerical aperture. For instance, fixing
$NA=0.5$, namely $\tilde\theta_i=\pi/6$, we find $\beta=0$ for $n=1$,
$\beta\approx 0.84$ for $n=2$ and $\beta=\pi/2$ for $n=3$ and $n=4$. The corresponding
values of the efficiency are quoted in section \ref{sec: apert} of the main text.
The value $\beta=0$ for $n=1$ could be expected since, for linear Thomson scattering, the maximum
of the scattered radiation is emitted perpendicular to the polarization direction. Nevertheless,
that is not the case for harmonic generation from nonlinear Thomson scattering.


\begin{thebibliography}{99}



\bibitem{CPA} D. Strickland and G. Mourou,  Opt. Commun. {\bf 56}, 219 (1985).

\bibitem{HERCULES2008} V. Yanovsky {\it et al}, Opt. Express {\bf16}, 2109 (2008).

\bibitem{mourou06}{G. A. Mourou, T. Tajima, and S. V. Bulanov, Rev. Mod. Phys. {\bf 78}, 310 (2006).}

\bibitem{qvac-reviews}{M. Marklund, P. K. Shukla, Rev. Mod. Phys. {\bf 78}, 591 (2006); 
F. Ehlotzky, K. Krajewska, J. Z. Kaminski, Rep. Prog. Phys. {\bf 72}, 046401 (2009);
A. Di Piazza, C. Muller, K. Z. Hatsagortsyan, and C. H. Keitel, Rev. Mod. Phys. {\bf 84}, 1177 (2012).}

\bibitem{PPSVsearch}{
S. L. Adler, Ann. Phys. {\bf 67}, 599 (1971);
E. B. Aleksandrov, A. A. Anselm, A. N. Moskalev,
Sov. Phys. JETP {\bf 62}, 680 (1985);
Y. J. Ding, A. E. Kaplan, Phys. Rev. Lett. {\bf 63}, 2725 (1989);
F. Moulin, D. Bernard, Opt. Comm. {\bf 164}, 137 (1999);
D. Bernard {\it et al.}, Eur. Phys. J. D {\bf 10}, 141 (2000);
G. Brodin, M. Marklund, L. Stenflo, Phys. Rev. Lett. {\bf 87}, 171801 (2001);
G. Brodin  {\it et al.} Phys. Lett. A {\bf 306}, 206 (2003);
V. I. Denisov , I. V. Krivchenkov, N. V. Kravtsov, Phys. Rev. D {\bf 69}, 066008 (2004);
T. Heinzl {\it et al.}, Opt. Comm. {\bf 267}, 318 (2006);
A. Di Piazza, K.Z. Hatsagortsyan, C.H. Keitel, Phys. Rev. Lett. {\bf 97}, 083603 (2006);
A. Ferrando, H. Michinel, M. Seco, D. Tommasini, Phys. Rev. Lett., {\bf 99}, 150404 (2007);
D. Tommasini, A. Ferrando, H. Michinel, M. Seco, Phys. Rev. A {\bf 77}, 042101 (2008);
E. Lundstrom {\it et al.}, Phys. Rev. Lett. {\bf 96}, 083602 (2006);
M. Marklund, J. Lundin, Eur. Phys. J. D {\bf 55}, 319 (2009); 
B. King, A. Di Piazza, and C. H. Keitel, Nat. Photonics {\bf 4}, 92 (2010); 
M. Marklund, Nat. Photonics {\bf 4}, 72 (2010); 
D. Tommasini, and H. Michinel, Phys. Rev. A {\bf 82}, 011803R (2010); 
B. King, A. Di Piazza and C. H. Keitel, Phys. Rev. A {\bf 82}, 032114 (2010); 
H. Gies and L. Roessler, Phys. Rev. D {\bf 84}, 065035 (2011);
B. King, C. H. Keitel, New J. Phys. {\bf 14}, 103002 (2012);
Qiang-Lin Hu {\it et al.} Phys. of Plasmas {\bf 19}, 042306 (2012);
L. Kovachev, D. A. Georgieva, A. Daniela, K. L. Kovachev, Opt. Lett. {\bf 37}, 4047 (2012);
E. Milotti, {\it et al.}, Int. J. Quant. Inf. {\bf 10}, 1241002 (2012); 
J. K. Koga {\it et al.} Phys. Rev. A {\bf 86}, 053823 (2012);
R. Battesti, and C. Rizzo, Rep. Prog. Phys. {\bf 76}, 016401 (2013).
}


\bibitem{new_physics}{M. Bregant {\it et al.}, PVLAS, Phys. Rev. D {\bf 78}, 032006 (2008); 
G. Zavattini and E. Calloni, Eur. Phys. J. C {\bf 62}, 459 (2009); 
D. Tommasini, A. Ferrando, H. Michinel, M. Seco, J. High Energy Phys. {\bf 11} 043 (2009); 
B. Dobrich, H. Gies, J. High Energy Phys. {\bf 10}, 022 (2010); 
A. Accioly, P. Gaete, J. A. Helayel-Neto, Int. J. Mod. Phys. {\bf 25}, 5951 (2010);
B. Doebrich, A. Eichhorn, J. High Energy Phys. {\bf 06}, 156 (2012);
G. Zavattini, {\it et al.} Int. J Mod. Phys. {\bf 27}, 1260017 (2012);
% Magnetically amplified light-shining-through-walls via virtual minicharged particles
B. Dobrich, H. Gies, N. Neitz, F. Karbstein, Phys. Rev. D {\bf 87}, 025022 (2013);
% Exact self-similar solutions in Born-Infeld theory
E. Y. Petrov, A. V. Kudrin, Phys. Rev. D {\bf 87}, 087703 (2013).}


\bibitem{pressure} A. Paredes, D. Novoa and D. Tommasini, Phys. Rev. Lett. {\bf 109},
253903 (2012).

\bibitem{redhead98} P. A. Redhead, \emph{Ultrahigh and Extreme High Vacuum}, in "Foundations of Vacuum Science and Technology", p. 625, Ed. J. M. Lafferty 
(Wiley, New York) (1998).


\bibitem{redhead} P.A. Redhead, 
%{\it Extreme high vacuum}, 
Cern accelerator school vacuum technology, proceedings,
 Cern reports 99, 213-226 (1999).

\bibitem {calcatelli} A. Calcatelli, 
Measurement {\bf 46}-2, 1029-1039 (2013).

\bibitem{Chen87} J. Z. Chen, C. D. Suen and Y.H. Kuo, J. Vac. Sci. Technol. A. {\bf 5}, 2373 (1987).


\bibitem{haffner} H. H\"affner {\it et al.}, Eur. Phys. J. D {\bf 22}, 163-182 (2003).



\bibitem{NTS} 
 J.H. Eberly and A. Sleeper, Phys. Rev. {\bf 176}, 1570 (1968);
  E. Esarey, S.K. Ride, P. Sprangle, 
 %{\it Nonlinear Thomson scattering of intense laser pnlses from beams and plasmas}
 Phys. Rev. E {\bf 48}, 3003 (1993);
S.-Y. Chen, A. Maksimchuk, D. Umstadter, 
%{\it Experimental observation of relativistic nonlinear Thomson scattering}, 
Nature  {\bf  396}, 653 (1998);
S.Y. Chen, A. Maksimchuk, E. Esarey, D. Umstadter, 
%{\it Observation of Phase-Matched Relativistic Harmonic Generation},
 Phys. Rev. Lett., {\bf 84}, 5528 (2000);
 M. Babzien {\it et al.}, Phys. Rev. Lett. {\bf 96}, 054802(2006).
 T. Kumita {\it et al.} Laser Phys. {\bf 16}, 267 (2006).

\bibitem{sarachik}  E.S. Sarachik,  G.T. Schappert, 
%{\it Classical theory of the scattering ofintense laser radiation by free electrons},
 Phys. Rev. D  {\bf 1}, 2738 (1970). 

\bibitem{salamin}  Y.I. Salamin, F.H.M. Faisal, 
 %{\it Harmonic generation by superintense light scattering from relativistic electrons}, 
 Phys. Rev. A  {\bf 54}, 4383 (1996).
 
 \bibitem{salamin2}  Y.I. Salamin, F.H.M. Faisal, 
  Phys. Rev. A  {\bf 55}, 3694 (1997).

\bibitem{lawson} J. D. Lawson, IEEE Trans. Nucl. Sci. {\bf NS-26}, 4217 (1979);
E. Esarey, C. B. Schroeder, and W. P. Leemans, Rev. Mod. Phys. {\bf 81} ,1229 (2009).

\bibitem{barriersupH} N.B. Delone, V.P. Krainov, 
%{\it Tunneling and barrier-suppression ionization of atoms and ions in a a laser radiation field},
 Uspekhi Fizicheskikh Nauk {\bf 168} (5), 531 (1998);
V.V. Strelkov, A.F. Sterjantov, N.Yu Shubin, V.T. Platonenko, 
%{\it XUV generation with several-cycle laser pulse in barrier-suppression regime}, 
J. Phys. B: At. Mol. Opt. Phys. {\bf 39}, 577 (2006). 

\bibitem{gibbon} P. Gibbon, {\it Short Pulse Laser Interaction with Matter},
Imperial College Press  (2007).

\bibitem{detectors1} A. Korneev {\em et al.},
% Quantum efficiency and noise equivalente power of nanostructured NbN single-photon detectors in the wavelength range from visible to infrarred",
IEEE Trans. Appl. Supercon. {\bf 15}, 2, 571 (2005).

\bibitem{detectors2} K. Smirnov {\em et al.},
% Ultrathin NbN film superconducting single-photon detector array"
J. Phys.: Conf. Ser. {\bf 61}, 1081 (2007).

\bibitem{HRS} K. Clays, A. Persoons, Phys. Rev. Let. {\bf 66}, 2980 (1991);
K. Clays, A. Persoons, Rev., Sci. Instrum. {\bf 6} 3285 (1992); K. Clays, A. Persoons,
L. De Maeyer, Adv. Chem. Phys. {\bf 85} (III), 455 (1994).

\bibitem{HRS2} E. Hendrickx, K. Clays, A. Persoons, Acc. Chem. Res., {\bf 31}, 675
(1998).

\bibitem{bryant1965}
%Extreme Vacuum Technology Developments
P. J. Bryant {\it et al.}, NASA CR-324 (1965).


\bibitem{fernandez2012}
%A Large Hadron Electron Collider at CERN
J. L. Abelleira-Fern\'andez {\it et al.}, J. Phys. G:
Nucl. Part. Phys. 
{\bf 39} 075001 (2012).


\bibitem{VEGA} http://www.clpu.es/es/infraestructuras/linea-

principal/fase-3.html .

\bibitem{JuSPARC} Annual Report 2011 of the IKP, Berichte des 
Forschung\-szentrums Jülich
Juel-4349, ISSN: 0944-2952. 
http://

donald.cc.kfa-juelich.de/wochenplan/publications/

AR2011/documents/AR2011\_Highlights.pdf

\bibitem{JuSPARC2} M. B\"uscher, private communication.

\bibitem{ELI} http://www.extreme-light-infraestructure.eu.

\bibitem{eisaman} M.D. Eisaman, J. Fan, A. Migdall, S.V. Polyakov,
Rev. Sci. Instrum {\bf 82} 071101 (2011).


\bibitem{mackenroth} F. Mackenroth, A. Di Piazza, 
%{\it Nonlinear Compton scattering in ultrashort laser pulses}
Phys. Rev. A {\bf 83}, 032106 (2011).

\bibitem{krafft} G.A. Krafft, 
%{\it Spectral distributions of Thomson-scattered photons from high-intensity pulsed lasers}, 
Phys. Rev. Lett.  {\bf 92}, 204802 (2004).

\bibitem{Gao} J. Gao, 
%{\it Thomson scattering from ultrashort and ultraintense laser pulsed},
Phys. Rev. Lett.  {\bf 93}, 243001 (2004);
 Y. Tian {\it et al.}
%{\it Spectral distributions of harmonic generation from electron oscillation driven by
%intense femtosecond laser pulses},
 Opt. Comm.  {\bf 261}, 104 (2006);
 Y. Tian, Y. Zheng, Y. Lu, J. Yang, 
%{\it Electron dynamics and spatial characteristics
%of emission from electron oscillation driven by femtosecond laser pulses},
 Optik  {\bf 122}, 1373 (2011).

\bibitem{peak} 
O. Har-Shemesh, A. Di Piazza,
%{\it Peak intensity measurement of relativistic lasers via nonlinear Thomson scattering},
Opt. Lett.  {\bf 37}, 1352 (2012).

\bibitem{castillo} C.I. Castillo-Herrera, T.W. Johnston,
%{\it Incoherent harmonic emission from strong electromagnetic waves in plasmas}
IEEE Trans. Plasma Sci. {\bf 21}, 125 (1993).





\end{thebibliography}
\end{document}